\documentclass[prl,preprint,superscriptaddress]{revtex4}
\usepackage{latexsym}
\usepackage{graphicx}
\usepackage{epsfig}
\usepackage{amssymb}
\usepackage{amsfonts}
\usepackage{mathrsfs}
\usepackage{amsmath}

\begin{document}

\title{More than one dynamic crossover in protein hydration water}

\author{Marco G. Mazza}
\affiliation{Center for Polymer Studies and Department of Physics, Boston University -- Boston, MA 02215 USA}
\author{Kevin Stokely}
\affiliation{Center for Polymer Studies and Department of Physics, Boston University -- Boston, MA 02215 USA}
\author{Sara~E.~Pagnotta}
\affiliation{Centro de Fisica
    de Materiales (CSIC-UPV/EHU) - Materials Physics Center MPC,
    Donostia-San Sebastian, Spain.} 
\author{Fabio~Bruni}
\affiliation{Dipartimento
    di Fisica ``E.  Amaldi'', Universit\`a di Roma Tre, 00146 Roma,
    Italy.} 
\author{H.~Eugene Stanley}
\affiliation{Center for Polymer Studies and Department of Physics, Boston University -- Boston, MA 02215 USA}
\author{Giancarlo Franzese}
\affiliation{Departament de F\'{\i}sica Fonamental, Universitat de
    Barcelona, Diagonal 647, 08028 Barcelona, Spain.}

\begin{abstract}


  Studies of liquid water in its supercooled region have led to many insights
  into the structure and behavior of water.  While bulk water freezes at its
  homogeneous nucleation temperature of approximately $235$~K, for protein
  hydration water, the binding of water molecules to the protein avoids
  crystallization.  Here we study the dynamics of the hydrogen bond (HB)
  network of a percolating layer of water molecules, comparing measurements of
  a hydrated globular protein with the results of a coarse-grained
  model that has been shown to successfully reproduce the properties
  of hydration water.
With dielectric spectroscopy we measure the temperature
  dependence of the relaxation time of protons charge fluctuations. These
  fluctuations are associated to the dynamics of the HB network of water
  molecules adsorbed on the protein surface. With Monte Carlo (MC) simulations
  and mean--field (MF) calculations we study the dynamics and thermodynamics
  of the model.  In both experimental and model analyses we find two dynamic
  crossovers: (i) one at about $252$~K, and (ii) one at about $181$~K.  The
  agreement of the experiments with the model allows us to relate the two
  crossovers to the presence of two specific heat maxima at ambient pressure.
  The first is due to fluctuations in the HB formation, and the second, at
  lower temperature, is due to the cooperative reordering of the HB network.

\end{abstract}

\maketitle

Recent experiments study water in the first hydration shell of
globular proteins \cite{che06,paw08,kho08a,kho08b,vog08}.  Contrary to bulk
water, this water does not freeze until the temperature $T$ is well below
$235$~K \cite{rup91}, with possibly important implications for biological
function \cite{fra09}.  While quasi-elastic neutron scattering investigations
\cite{che06} and some molecular dynamics simulations
\cite{kumar-biomolec,Lagi08} support the presence of a dynamic crossover at
about $220$~K, other experiments and simulations
\cite{paw08,kho08a,kho08b,vog09} do not.  It has been shown that the suggested
crossover could be related to the anomalous behavior of water, but independent
of a possible liquid--liquid critical point at finite $T$ \cite{KFS1}.

Here we show by experiments, simulations, and model calculations that the
dynamical properties of the HB network at the protein--water interface exhibit
not one, but two dynamic crossovers in the one-phase region at low
pressure. We show how the two crossovers are related to the
thermodynamics of water.  We investigate the dielectric relaxation time of
water protons, due to charged defects---such as H$_{3}$O$^{+}$---moving with a
diffusive or hopping mechanism 
along the HB network \cite{rup91,capaccioli98}. These measurements are a sensitive probe
for the HB breaking and formation~\cite{pey01}. We perform dielectric
relaxation experiments on lysozyme powder with hydration level $h=0.30$~g
H$_2$O/g dry protein, over a broad frequency ($10^{-2}$~s$^{-1} -
10^{8}$~s$^{-1}$) and temperature range ($150$~K$\leq T\leq 300$~K).  The
experimental set-up and the data
analysis~\cite{bru04,pag05,pag09,piz01,piz01b} are described in the Methods
section and Supporting Information.

\begin{figure}
\centering
\includegraphics[scale=0.48]{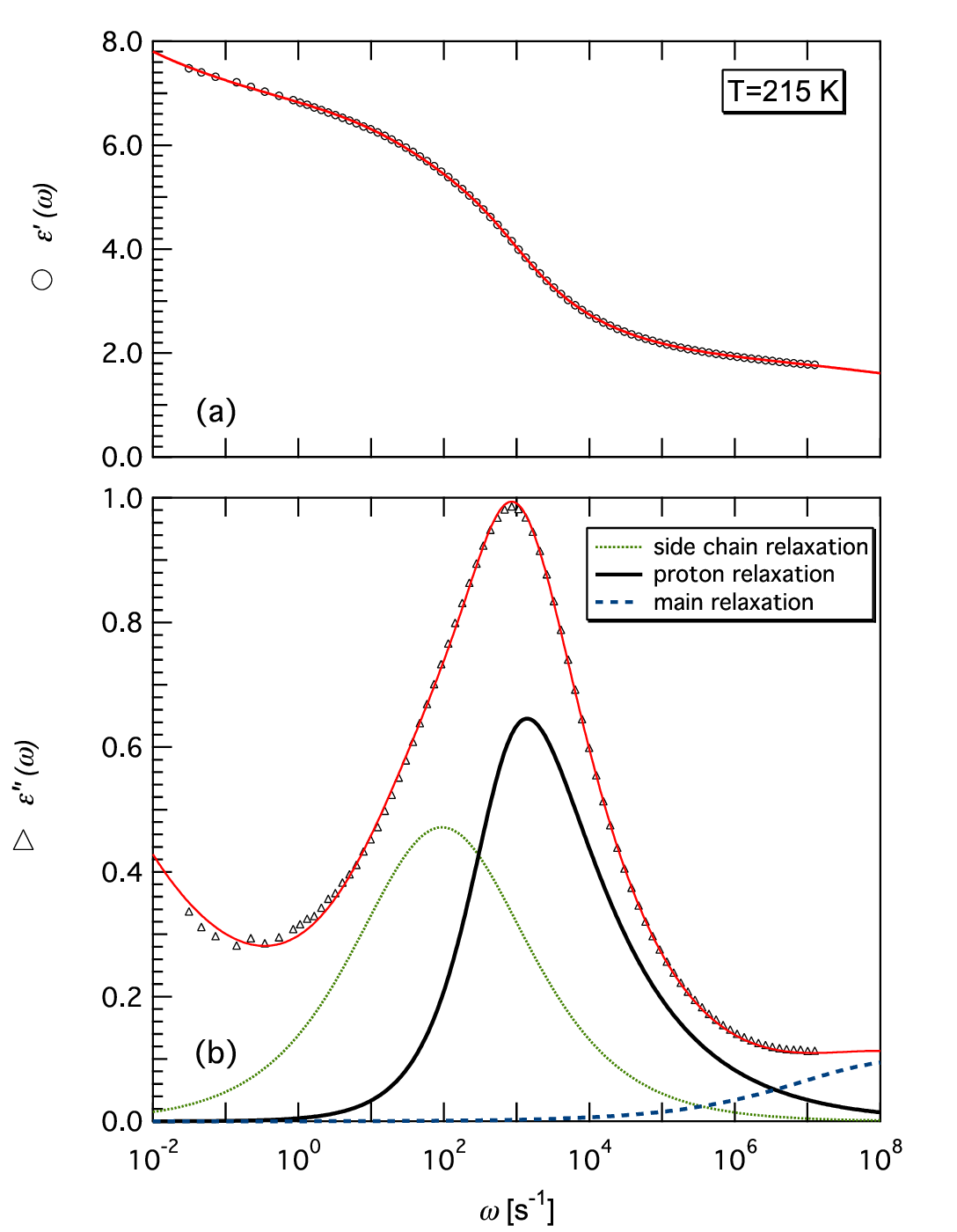}
\caption{Dielectric relaxation data for lysozyme powder at $215$~K and
  hydration level $h=0.30$~g H$_2$O/g dry protein. {\bf (a)}
  $\epsilon^{\prime}$ ($\circ$) and {\bf (b)}
  $\epsilon^{\prime\prime}$ ($\triangle$) are real and imaginary components,
  respectively, of the complex permittivity $\epsilon_{m}^{\ast} \equiv
  \epsilon^{\prime} -\imath\epsilon^{\prime\prime}$, and $\omega$ is the
  angular frequency.  Solid lines through symbols result from the fitting
  procedure in the complex plane
  \cite{bru04,pag05,pag09,piz01,piz01b}, described in the Method
  section and Supporting Information.
  As shown in (b) for $\epsilon^{\prime\prime}$, $\epsilon_{m}^{\ast}$
  is resolved into (i) a low-$\omega$ tail  
(omitted for  clarity), 
(ii) a broad process at intermediate $\omega$ deconvoluted into two
relaxations (continuous and dotted lines),  and (iii) a high-$\omega$  
relaxation  (dashed line). \label{spectra}}
\end{figure}

In the dielectric spectrum for lysozyme powder at 215 K,
(Fig.~\ref{spectra}a), we identify (i) a low frequency tail, (ii) a relatively broad
process at intermediate frequency, and (iii) a small high frequency relaxation. The
low frequency tail (i) is due to electrode polarization, to interfacial
dispersion--also known as Maxwell-Wagner effect--and to sample conductivity
(see Methods section). The high-frequency process (iii) has a relaxation time with a
$T$--dependence and absolute values identical to recent dielectric
measurements on the same protein at the same water content
$h$~\cite{kho08a,kho08b}: this process is labeled as ``main'' in the
above cited 
literature, and we will adopt this choice (see Methods section and
Fig.~1 in the Supporting Information). 
  This relaxation becomes undetectable at hydration
levels below $h\sim0.3$ g H$_2$O/g dry protein, and it has been assigned to a local relaxation
of protein hydration water \cite{kho08a}.

\begin{figure}
\centering
\includegraphics[scale=0.45]{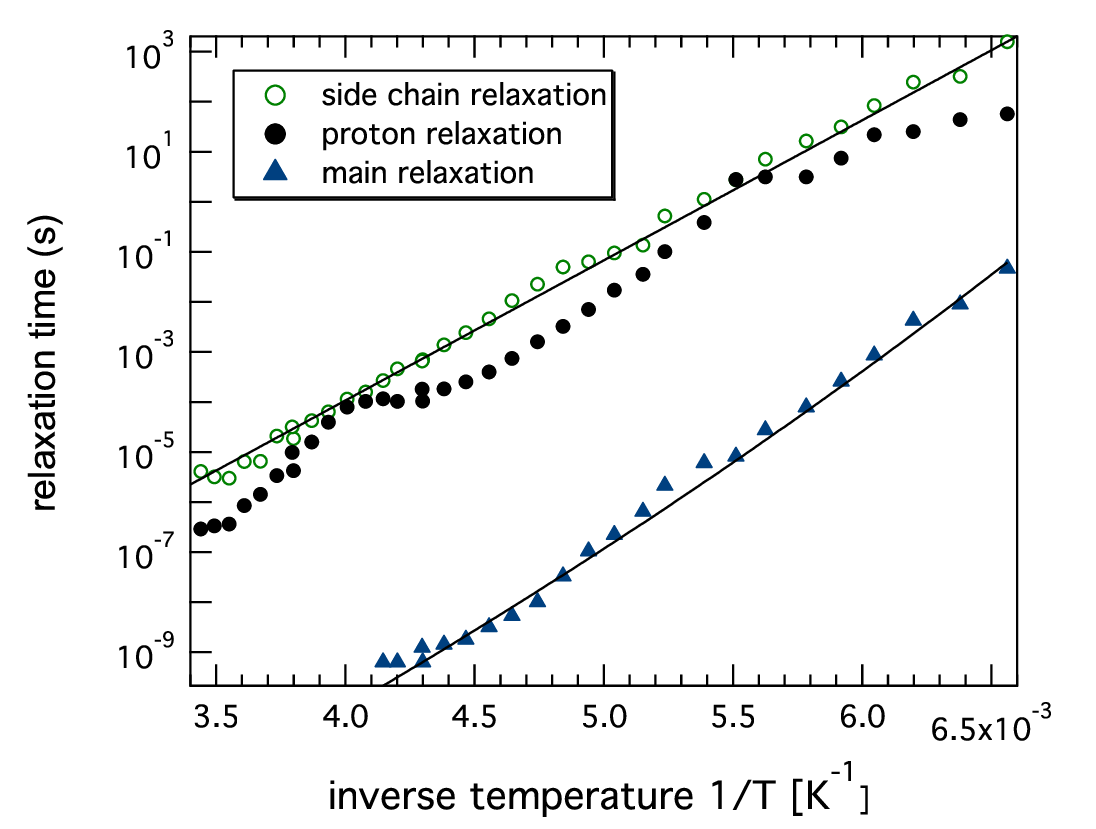}
\caption{Temperature dependence of the characteristic 
  times of the relaxation  processes
  shown in  Fig.~\ref{spectra}: 
``side chain'' relaxation ($\circ$) and 
``proton'' relaxation ($\bullet$) of  the
intermediate-$\omega$ relaxation (ii), and ``main'' relaxation
($\blacktriangle$) of 
the high-$\omega$  relaxation (iii).
Solid lines are Arrhenius fits. \label{Tdep}}
\end{figure}

\begin{figure}
\centering 
\includegraphics[scale=0.7]{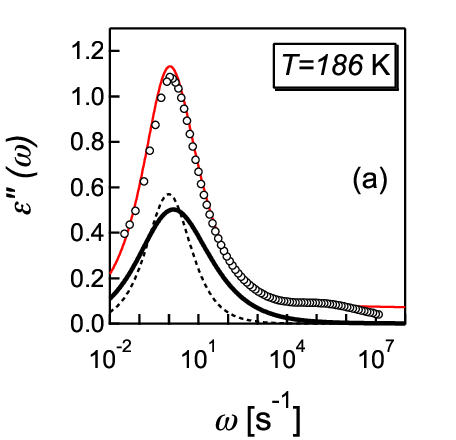}

\includegraphics[scale=0.7]{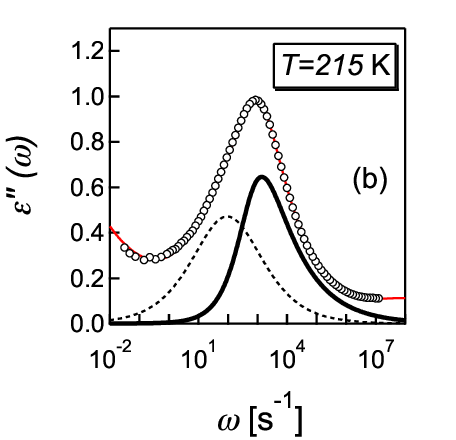}

\includegraphics[scale=0.7]{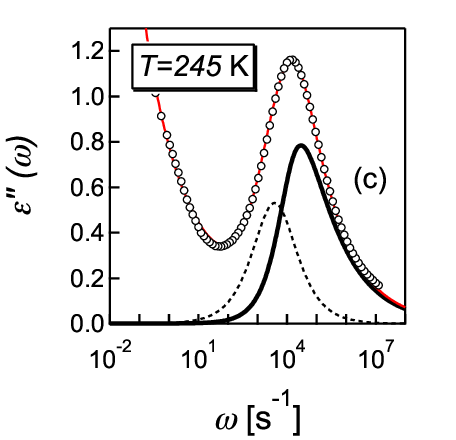}
\caption{Imaginary component $\epsilon^{\prime\prime} (\omega)$ of the
  measured complex permittivity as a function the angular frequency
  $\omega$, at three temperatures: 
{\bf (a)} $T=186$ K, 
{\bf (b)} $T=215$ K, 
{\bf (c)} $T=245$ K. 
Solid lines trough symbols are result
  of the fitting procedure in the complex plane (see Methods section and Supporting Information).
  Here we show the results of the decomposition into two processes of
  the broad relaxation peak. Dashed lines represent the
  ``side chain'' relaxation, while thick solid lines represents the
  ``proton'' relaxation.  The fitting procedure yields a $\beta_1
  \approx 1$ (see Equation 5) for the ``side chain'' relaxation over the
  entire temperature range investigated, resulting in a symmetric
  relaxation process. 
  Conversely, the same procedure gives a $\beta_2 < 1$ for the ``proton'' relaxation, resulting in a characteristic asymmetric shape. 
  We omit for clarity the contributions
  due to sample conductivity and to electrode polarization at low
  frequency. \label{panel}}
\end{figure}

The broad relaxation process (ii), whose width is about 3.5 frequency decades at half
maximum, can be resolved into two contributions
\cite{bru04} relatively close in frequency and largely overlapping but with a
resulting markedly different $T$-dependence (Fig.~\ref{Tdep}).  Such
decomposition has not been discussed in previous work
\cite{kho08a,kho08b,fra09}, but a quantitative test for the presence of two
relaxation processes has been described in detail in Ref.  \cite{bru04} (see
Methods section and Supporting Information).  The quality of the two-relaxation fit, along with the
$T$-dependence, $h$-dependence, and shape of each relaxation, impels 
us to
conclude that two separate relaxations are present (see Supporting Information). Here we label these processes ``side chain''
relaxation and ``proton'' relaxation (Fig.~\ref{spectra}).
This labeling is
based on previous studies on the $T$-dependence and $h$-dependence
of the
dielectric response of the same protein and of a similar globular protein
(myoglobin). In particular, the ``side chain'' process has a relaxation time whose
$T$-dependence and absolute values agree with those measured by others for
hydrated lysozyme powders \cite{kho08a} and hydrated myoglobin \cite{fra09}
(see Methods and Fig.~1 in the Supporting Information). 
The ``side chain'' relaxation has a symmetric shape over the entire temperature
range investigated (see Fig.  \ref{panel}), in agreement with previous
findings \cite{kho08a}. This observation provides additional support to
distinguish the ``side chain'' relaxation from the more asymmetric ``proton''
relaxation. Moreover, the wide temperature and frequency ranges investigated
allow to follow these two processes carefully and to identify them even when
they are largely overlapping. Fitting parameters for each process, such as its
relaxation time and shape parameters, change gradually with temperature, in
such a manner to make the distinction between ``side chain'' and ``proton'' relaxation
reliable at all temperatures.

The ``proton'' relaxation, contributing to the broad peak in Fig.~\ref{spectra},
is the object of the present report and has been extensively studied
\cite{rup91,bru04,pag05,pag09,piz01,piz01b} at hydration $h<0.30$ g H$_2$O/g dry protein.  At
these hydration levels we find that the ``side chain'' relaxation is indeed quite
small and shows an extremely strong slowing down with dehydration, in
agreement with Ref. \cite{kho08a}. The measured dielectric spectra for
lysozyme at $h<0.30$ g H$_2$O/g dry protein is therefore dominated by the ``proton'' relaxation
\cite{rup91,bru04,pag05,pag09,piz01,piz01b}. Its characteristic relaxation
time and the related dc conductivity has been described in terms of
percolation theory \cite{rup91,pag05}. Its assignment to water-assisted proton
displacements over the protein surface has been tested by measuring
hydrogenated and deuterated samples \cite{rup91,pag09}, and its quantum
character has been checked experimentally by dielectric spectroscopy and deep
inelastic neutron scattering \cite{pag09,sen07}, and verified against
theoretical models \cite{tuc01}.

\begin{figure}
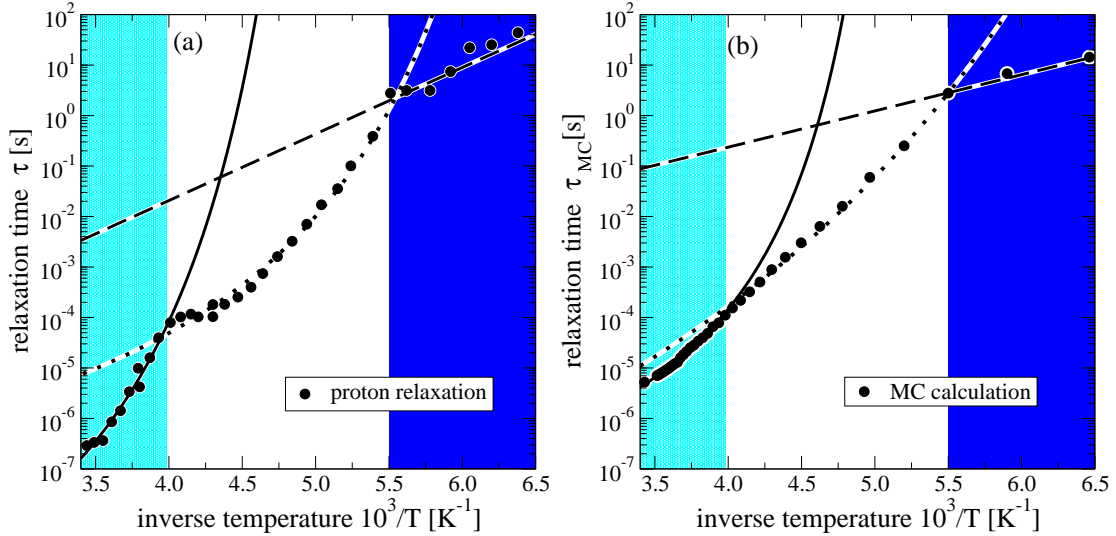

\centering 
\includegraphics[scale=0.4]{fig4a.eps}
\includegraphics[scale=0.4]{fig4b.eps}
\caption{Two crossovers in the ``proton'' relaxation time $\tau$ of
  hydration water. We find a non-Arrhenius to non-Arrhenius crossover at
  $T\approx 252$~K and a non-Arrhenius to Arrhenius crossover at
  $T\approx 181$~K.  {\bf (a)} Experimental $\tau$ ($\bullet$) vs.  $1/T$.
  Solid line is the VFT function with fitting parameters
  $\tau_{0}=7.8\times10^{-12}$~s, $B_T=9.4$~kJ/mol, $T_0=180$~K.  Dotted
  line is the VFT function with $\tau_{0}=6.5\times10^{-8}$~s,
  $B_T=6.2$~kJ/mol, $T_0=140$~K.  Dashed line is the Arrhenius function
  with fitting parameters $\tau_{0}=1.1\times10^{-7}$~s,
  $A=25.2$~kJ/mol.  The behaviors at high and low $T$ intersect at about
  $232$~K.
  {\bf (b)} MC relaxation time $\tau_{MC}$ ($\bullet$) vs.  $1/T$, for $P=0.1$~MPa.
Solid line is the VFT function with 
$\tau_{0}=1.61\times10^{-8}$~s, $B_T=5.2$~kJ/mol, $T_0=181.2$~K.  Dotted line
is the VFT function with 
$\tau_{0}=7.5\times10^{-10}$~s,
$B_T=15.9$~kJ/mol, $T_0=95.2$~K.  Dashed line is the Arrhenius function with
$\tau_{0}=3.3\times10^{-4}$~s, $A=13.7$~kJ/mol.
See the text for the discussion about the quantitative differences
between the numerical and the experimental results. \label{exp-crossover}}
\end{figure}

At hydration $h<0.30$ g H$_2$O/g dry protein, the proton relaxation process has a characteristic
relaxation time whose temperature dependence is well described by the
Vogel-Fulcher-Tamman (VFT) equation
$\tau(T)=\tau_0\exp\left[B_T/R(T-T_0)\right]$, where $\tau_0$, $B_T$ and $T_0$
are fitting parameters and $R$ is the gas constant.  For the sample water
content, $h=0.30$ g H$_2$O/g dry protein, the proton relaxation time $\tau$ at high $T$ shows a
VFT behavior, with a clear kink at $T\approx252$~K
(Fig.~\ref{exp-crossover}a), where we find a crossover from one high-$T$ VFT
behavior to a second VFT behavior at lower temperature.  We associate the
crossover at $\approx252$~K to a change of the diffusion regime of water
protons, from sub-diffusive at lower $T$ to freely diffusive at higher $T$,
that has been previously reported for a lysozyme sample at $h=0.26$ g H$_2$O/g dry protein and
$260$ K \cite{pag05}.  Interestingly, the VFT description at high temperature
of the water proton relaxation time is characterized by the same $T_{0} \sim
200$~K independent of sample hydration \cite{bru04,pag05,pag09}.  Moreover,
$\tau$ would reach $100$ s at $\sim220$~K (Fig.~\ref{exp-crossover}a): this is
often defined as the dielectric glass transition temperature. This value is in
good agreement with the calorimetric glass transition temperature of the
hydrated protein at the same hydration level~\cite{kho08a,kho08b,pey01,pag05,piz01,piz01b}.

At $T\approx181$~K we observe a second crossover, as the $T$--dependence of
$\tau$ changes from the VFT above $181$~K to Arrhenius
$\tau(T)=\tau_0\exp\left(A/RT\right)$ below $181$~K, where $\tau_0$ is a
characteristic relaxation time and $A$ is a constant activation energy
(Fig.~\ref{exp-crossover}a).  The crossover at $\approx181$ K, to our
knowledge, has never been reported before. Here we offer its possible
interpretation based on the results of the model described below: we assign
this crossover to a structural rearrangement of the HB network.

It must be noted that neutron scattering experiments on the same hydrated
protein also revealed two dynamical transitions, one at $220$ K and the other
at $150$ K~\cite{zan08}. These transitions have been attributed respectively
to the rotational motion of interfacial water, and to proton dynamics on a
local (few \AA) scale. In particular, the low temperature transition ($150$~K)
was claimed to be due to a sudden increase of the configurational entropy of
the system, linked to a significant excursion of the HB length~\cite{zan08}.
However, deep inelastic neutron scattering experiments on the same hydrated
protein do not support this claim, as no change of the HB length and of the
proton potential can be measured up to $290$ K~\cite{pag09,sen07}.

We understand the microscopic mechanisms responsible for the dynamic
crossovers observed at $252$~K and $181$~K, by means of a coarse-grained model
of a monolayer of water adsorbed on a generic inert substrate, representing
the low-hydrated protein powder.  At the hydration level considered here, 
the adsorption is such that water molecules
are restricted to a surface geometry with coordination number up to
four~\cite{Marzel02}, and the temperature of their structural
arrest  is expected to be much higher than $250$~K~\cite{Miyazaki}, hence
they  do not diffuse. Yet water molecules
do not crystallize~\cite{Miyazaki}, therefore, their positions and
orientations fluctuate, forming and breaking HBs. 
These assumptions are
grounded on the observation that, at the relatively low $h$ value
investigated, we have less than one monolayer of water molecules covering the
protein surface, and the protein itself does not undergo any configurational
transformation or large scale motion~\cite{rup91,Miyazaki}.  This model,
originally proposed in Ref.~\cite{franzese} and extensively studied,
e. g.,  in
Refs.~\cite{KFS1,franzese3,franzese2,KFS2008,MazzaCPC09,Stokely,Strekalova},
reproduces the known properties of water at interfaces, including the shape of the
locus of temperatures of maximum density in the $(T,P)$ plane, the
anomalous behavior of thermodynamic response functions, the
 subdiffusive regime at low $T$ for protein hydration water,  the 
occurrence of minima and maxima in diffusivity upon
pressurization~\cite{GFcondmat}, and has predicted behaviors for
protein hydration water,
successively verified by experiments~\cite{KFS1,F2008}.

The model discretizes the coordinates of the water molecules in the monolayer into $N$
cells, each containing one molecule, with a volume given by the inverse of the
average water density and a height given by the monolayer thickness. To take
into account the change of configurational entropy upon HB formation,
the model
associates to each molecule $i$ a variable $\sigma_{ij}=1,2,...,q$ with a
discrete number of states $q$ describing the bonding state with a neighbor
molecule $j$. The model chooses $q$ by adopting the standard convention that $30^o$ is
the maximum deviation of a HB from a linear bond, i.e. $q\equiv 180^o/30^o=6$.
For every molecule there are $q^4=6^4\equiv 1296$ total possible bonding
states.  The system is fully specified by the average density $V/N$ and the
set of $\sigma_{ij}$.

The model separates the interactions among molecules into three
components.  The first is the sum of all 
isotropic interactions (e.g., van der Waals) between molecules at distance
$r\equiv(V/N)^{1/d}$, and is represented by a Lennard--Jones potential,
$U_0(r)=\epsilon[(r/r_0)^{12}-(r/r_0)^6]$.  On the basis of previous
experiments, the model choose attractive energy
$\epsilon=5.8$~kJ/mol~\cite{Henry2002} and
$r_0\equiv(v_0)^{1/d}=2.9$~\AA~\cite{Narten67}.

The second is the directional component of the HB interaction.  Neighboring
molecules $i$ and $j$ form a HB when their facing bonding variables are in the
same state, i.e. $\sigma_{ij} = \sigma_{ji}$.  Formation of a bond leads to a
decrease of local energy by amount $J=2.9$ kJ/mol, and an increase of local
volume by amount $v_{\rm HB}$. Based on the change of density between
tetrahedral ice Ih and interpenetrating tetrahedral network as in ice VI or
ice VIII, the model choose $v_{HB}/v_0=0.5$. Note that the volume increase $v_{HB}$
does not correspond to a larger separation between molecules, but to a larger
volume per molecule, due to a decrease of number of neighbors. As a
consequence, the increase $v_{HB}$ does not affect the $U_0(r)$ term.  The
total contribution to the enthalpy is given by $-(J-Pv_{\rm HB})N_{\rm HB}$,
where $P$ is pressure, and $N_{\rm HB}$ is the number of HBs in the
system. $N_{\rm HB}$ is a function of the configuration of variables
$\sigma_{ij}$
\begin{equation}
N_{\rm HB}=\sum_{\langle i,j \rangle}\delta_{\sigma_{ij},\sigma_{ji}}
\end{equation}
where $\langle i,j \rangle$ indicates nearest--neighbors, and $\delta_{a,b}=1$ if $a=b$,
else $\delta_{a,b}=0$.

The final interaction is the cooperative (i. e. many--body) interaction among
HBs which gives rise to O--O--O correlation~\cite{Ricci}, locally driving the
molecules toward an ordered configuration.  This is modeled with an
interaction among the four $\sigma_{ij}$ within the same cell, driving them
towards the same state.  There is a local energy decrease by an amount
$J_\sigma=0.29$ kJ/mol for each of the six possible pairs of $\sigma_{ij}$
within the same cell which are in the same state.

The enthalpy of this model at pressure P is~\cite{franzese,franzese2,franzese3}
\begin{equation}
H= \sum_{ij} U_0(r_{ij}) -(J -Pv_{\rm HB})N_{\rm HB}
-J_\sigma \sum_{(k,l)_i} \delta_{\sigma_{ik},\sigma_{il}}
+PV_0,
\label{hamilt}
\end{equation}
\noindent
where the first sum is over all pairs of molecules $(i,j)$, the second sum is
over all pairs of $\sigma_{ik}$ belonging to molecule $i$, and $V_0\geqslant N
v_0$ is the water volume without counting the contribution of the HBs.

We perform MC simulations at constant $N=10^4$, $P$ and $T$, where $V_0$
fluctuates and the configuration of variables $\sigma_{ij}$, and therefore
$N_{\rm HB}$, changes, resulting in a variable total volume $V\equiv V_0+
N_{\rm HB} v_{\rm HB}$. We explore the thermodynamics of the system in
the $P$--$T$ plane, confirming previous results
\cite{KFS1,franzese,franzese3,franzese2,KFS2008,MazzaCPC09,Stokely,nota_MC}.

Next, we study the dynamic evolution of the system, by adopting the single-spin flip
implemented through the Metropolis algorithm, corresponding to Model A in the
classification of Hohenberg and Halperin~\cite{hohenberg77}.  The most
straightforward quantity to compare with the experiments considered here is
the autocorrelation function
\begin{equation}
C_M(t)\equiv\frac{1}{N}\sum_{i=1}^N \frac{\langle
  M_i(t_0+t)M_i(t_0)\rangle-\langle M_i \rangle^2} {\langle M_i(t_0)^2
  \rangle-\langle M_i \rangle^2}
\label{corr}
\end{equation}
where $t$ is the time measured in MC steps, $t_0$ is a time larger than the
equilibration time of the system, $M_i\equiv\frac{1}{4}\sum_j\sigma_{ij}$
quantifies the order of the four bond indices $\sigma_{ij}$ of molecule $i$.
Defining the MC relaxation time $\tau_{\rm MC}$ from $C_M(\tau_{\rm MC})=1/e$,
we observe two crossovers in the computed $\tau_{\rm MC}$: from VFT to VFT at
higher $T$, and from VFT to Arrhenius at lower $T$.  We rescale our
experimental data using real units, setting the higher--$T$ crossover at
$T\approx 252$~K \cite{KFS} and the lower--$T$ crossover at $T\approx 181$~K
(Fig.~\ref{exp-crossover}b) \cite{nota_MC}.
We find a good agreement of the MC results with the experimental data in
Fig.~\ref{exp-crossover}, with some difference at high $T$ that we will
discuss in the following.  The agreement suggests that the model could provide
a good description of the dynamics and connectivity of the real HB network,
thus allowing us to use it as a tool to investigate the thermodynamic origin
of the observed crossovers.

\begin{figure}
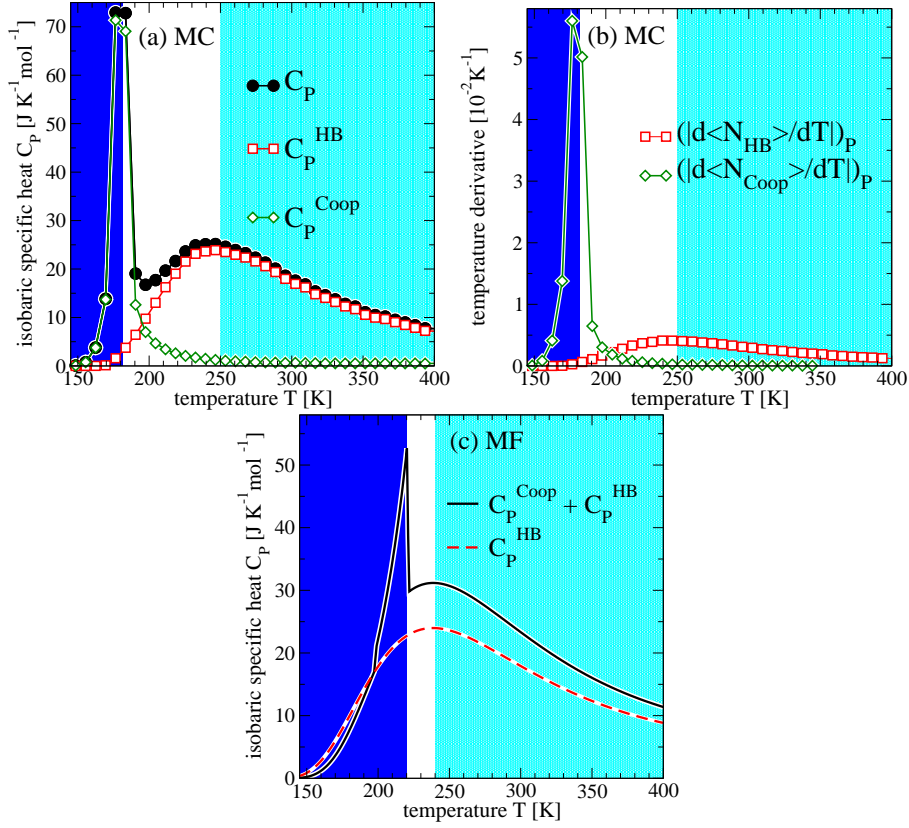

\centering 
\includegraphics[scale=0.32]{fig5a.eps}
\includegraphics[scale=0.32]{fig5b.eps}
\includegraphics[scale=0.32]{fig5c.eps}
\caption{Two maxima in the specific heat for the model.
{\bf (a)} The MC isobaric specific heat $C_{\rm P}$ ($\bullet$), at
  $P=0.1$~MPa has two maxima, decomposed into the components $C_{\rm
    P}^{{\rm HB}}$ ($\square$) and $C_{\rm P}^{{\rm Coop}}$ ($\diamond$)
  described in the text.  
{\bf (b)} $|{\rm d} \langle N_{\rm HB}\rangle/{\rm d} T|_{\rm P}$
($\square$) and $|{\rm d} 
\langle N_{\rm Coop}\rangle /{\rm d} T|_{\rm P}$ ($\diamond$) show
maxima where $C_{\rm 
P}^{{\rm HB}}$ and $C_{\rm P}^{{\rm Coop}}$, respectively, have maxima.
{\bf (c)} MF calculations at $P=0$ for the case with cooperative interaction
($J_\sigma/\epsilon=0.05$, continuous line labeled as $C_{\rm P}^{\rm
  Coop}+C_{\rm P}^{\rm HB}$) show two maxima for $C_{\rm P}$, while for the
case without cooperative interaction ($J_\sigma=0$, dashed line labeled as
$C_{\rm P}^{{\rm HB}}$) there is no low--$T$ maximum that, hence, is due to
the cooperativity. \label{cP}}
\end{figure}

Next, we discuss the thermodynamic interpretation of the crossover in the
model.  Reference~\cite{KFS1} shows by direct calculations of the model in
Eq.(\ref{hamilt}) that a maximum in isobaric specific heat $C_{\rm
  P}(T)\equiv\left(\partial H/\partial T\right)_{\rm P}$, implies a crossover
in the temperature dependence of $\tau$. This result is consistent with the
Adam--Gibbs theory~\cite{ag}. In the present work we find the $C_{\rm P}$
maximum observed in Ref.~\cite{KFS1}, and also another maximum at a lower $T$,
in a region not explored in Ref.~\cite{KFS1} (Fig.~\ref{cP}a). 
To understand the origin of the two $C_{\rm P}$ maxima, we write
the enthalpy as the sum of two terms $H=H^{{\rm HB}}+H^{{\rm Coop}}$, where
$H^{\rm HB}\equiv\langle-(J-Pv_{\rm HB}) N_{\rm HB}+ PV_0\rangle$, and
$H^{{\rm Coop}}\equiv H-H^{\rm HB}$.  We define the HB contribution to the
specific heat $C_{\rm P}^{{\rm HB}}\equiv\left(\partial H^{{\rm HB}}/\partial
  T\right)_{\rm P}$, and the cooperative contribution $C_{\rm P}^{{\rm
    Coop}}\equiv\left(\partial H^{{\rm Coop}}/\partial T\right)_{\rm P}$.
$C_{\rm P}^{{\rm HB}}$ is responsible for the broad maximum at higher $T$
(Fig.~\ref{cP}a).  To show that $C_{\rm P}^{{\rm HB}}$ captures the enthalpy
fluctuations due to the HB formation, we calculate the locus of maximum
fluctuation of $\langle N_{\rm HB} \rangle$, related to the maximum of $|{\rm
  d}\langle N_{\rm HB}\rangle /{\rm d}T|_{\rm P}$.  The temperatures of these
maxima coincide with the locus of maxima of $C_{\rm P}^{{\rm HB}}$
(Fig.~\ref{cP}b).

The maximum of $C_{\rm P}$ at lower $T$ is given by the maximum of $C_{\rm
  P}^{{\rm Coop}}$ (Fig.~\ref{cP}a).  To confirm that $C_{\rm P}^{{\rm
    Coop}}$ corresponds to the enthalpy fluctuations due to the cooperative
$J_\sigma$-term in Eq.(\ref{hamilt}), we calculate $|{\rm d}\langle N_{\rm
  Coop}\rangle/{\rm d}T|_{\rm P}$, where $\langle N_{\rm Coop}\rangle$ is the
average number of molecules with perfect local order of their bond indices.
We find that the locus of maxima of $|{\rm d}N_{\rm Coop}/{\rm d}T|_{\rm P}$
overlaps with the locus of maxima of $C_{\rm P}^{{\rm Coop}}$
(Fig.~\ref{cP}b). The same qualitative behavior for $C_{\rm P}$ is predicted
from MF calculations \cite{Stokely} for the cell model (Fig.~\ref{cP}c).

The non-monotonic behavior of $\langle N_{\rm HB}\rangle$ and $\langle N_{\rm
  Coop}\rangle$ explains the two crossovers in the HB correlation time. At
very low $T$, both experimental $\tau$ and simulation $\tau_{\rm MC}$ have an
Arrhenius behavior with constant activation energy $A$: $25.2$~kJ/mol in the
experiments and $13.7$~kJ/mol in the model. The quantitative difference
between the two arises from the choice of the parameters $J$ and $J_\sigma$.
In both experiments and model, $A$ is consistent with the average energy
$\langle E_{\rm HB} \rangle$ necessary to break a HB in a locally ordered
environment. The relation $A\approx \langle E_{\rm HB} \rangle$ in both
experiments and model suggests that the dynamics is dominated by the breaking
and formation of a single HB at low $T$. This is well understood in the model
where the energies $A\approx \langle E_{\rm HB} \rangle$ are both functions of
$\langle N_{\rm HB}\rangle$ and $\langle N_{\rm Coop}\rangle$ \cite{KFS1}.
Therefore, the saturation of the HB network ($|{\rm d}\langle N_{\rm
  HB}\rangle/{\rm d}T|_{\rm P}\approx 0$) and its ordering ($|{\rm d}\langle
N_{\rm Coop}\rangle/{\rm d}T|_{\rm P}\approx 0$) at low $T$ imply constant $A$
and an Arrhenius behavior for the HB correlation time.

At high $T$ where $C_P$ is monotonic, $\langle N_{\rm HB}\rangle$ and $\langle
N_{\rm Coop}\rangle$ increase for decreasing $T$. Hence, the activation energy
and $\langle E_{\rm HB} \rangle$ also increase, implying a non-Arrhenius
behavior.

At intermediate $T$, between the two maxima of $C_P$, the rate of change of
$\langle E_{\rm HB}\rangle$ is proportional to the decreasing $|{\rm d}\langle
N_{\rm HB}\rangle /{\rm d}T|_{\rm P}$ and the increasing $|{\rm d}\langle
N_{\rm Coop}\rangle /{\rm d}T|_{\rm P}$, giving rise to another non-Arrhenius
behavior down to the temperature of the maximum $|{\rm d}\langle N_{\rm
  Coop}\rangle /{\rm d}T|_{\rm P}$ and the crossover to Arrhenius
behavior~\cite{noteNgai2}. The difficulty to separate the large 
lysozyme contribution and the low-$h$ water contribution from 
the total experimental $C_P$  makes not possible a straightforward comparison
of our $C_P$ calculations with experimental data~\cite{zan08,Miyazaki}.

The relaxation time calculated for the model is characteristic to the breaking
and forming of H bonds, which is analogous to the proton relaxation measured
by dielectric spectroscopy.  We find good qualitative agreement between $\tau$
and $\tau_{\rm MC}$, but at high $T$ the crossover for $\tau$ is more
pronounced than that for $\tau_{\rm MC}$ (Fig.~\ref{exp-crossover}).  This
difference arises from two factors.

(i) The experiments are carried out at constant $h$, corresponding to a
decreasing effective $P$ (possibly negative due to the surface adsorption)
acting on water for decreasing $T$, while the MC results are at constant
$P=0.1$~MPa.  Our MF calculations predict that $C_{\rm P}$ displays two maxima
along any path $P(T)\lesssim 0.1$~MPa.
Along a path such as in the experiments, in which $P(T)$ decreases
monotonically upon cooling, $\langle E_{\rm HB}\rangle$ increases more rapidly
by decreasing $T$, because $\langle N_{\rm HB} \rangle$ and $\langle N_{\rm
  Coop} \rangle$ increase more rapidly when both $P$ and $T$ decrease
\cite{franzese2,KFS2008}. This allows $\tau_{\rm MC}$ to converge to the
experimental $\tau$ at high $T$.

(ii) The fluctuations in the HB network and distance between water oxygens,
predicted by the model, could enhance the probability for a proton to be
delocalized between two first-neighbor oxygens, inducing shorter proton
relaxation times than those predicted on the base of classical simulations at
high $T$.  Experiments~\cite{pag09} show that this effect is maximum around
$250$ K, approximately where the model predicts the maximum fluctuation of the
HB network and the experimental $\tau$ shows a stronger cusp than $\tau_{\rm
  MC}$.

To conclude, in dielectric spectroscopy experiments on hydrated lysozyme at
low hydration level, we observe a relaxation mode associated to water protons,
with two crossovers: one at $\approx 252$~K and another at $\approx 181$~K.
At the same time we find that a coarse-grained model of an adsorbed monolayer
of water shows in simulations two crossovers for the HB dynamics.  In the
model these two crossovers can be fully understood as the effects of two
structural changes of the HB network.  These two structural reorganizations
are marked by two maxima in $C_P$, as well as in the compressibility $K_T$ and
the isobaric thermal expansion coefficient $\alpha_P$ (not shown here).  The
two structural changes are (i) at higher $T$, associated with the maximum
fluctuations of the formation and breaking of the HBs, and (ii) at lower $T$,
associated with the maximum fluctuation of the ordering of the local
arrangement of the HBs. We argue that the model predictions provide an
interpretation for our experimental findings.

\section{Methods}

\subsection{Experimental set-up and data analysis}

We use an Alpha Analyzer dielectric apparatus (Novocontrol) to study
crystallized and highly purified lysozyme powder from chicken egg
white (Sigma-Aldrich), dialyzed and lyophilized to set its pH, then
re-hydrated \cite{bru04}.  The capacitor containing the sample has
blocking electrodes, coated by Teflon with thickness $\sim 1/40$ of
the sample thickness. This choice of thickness eliminates the
possibility of artifacts in our raw dielectric data, as
discussed in a recent publication \cite{ric09}, and provides a ``first
check'' of the reliability of the data analysis procedure adopted here.

Our measured complex admittance $Y_{m}(\omega)$ is directly related to
the complex permittivity
$\epsilon_{m}^{\ast}(\omega)=\epsilon_{m}^{\prime}(\omega) -
\jmath\epsilon_{m}^{\prime\prime}(\omega)$ given that

\begin{equation}
\epsilon_{m}^{\ast}(\omega)=\frac{h}{\imath\omega\epsilon_{o}S}Y_{m}(\omega).
\end{equation}
Here $\imath = \sqrt{-1}$, $\epsilon_{o}$ is the permittivity of free
space, $S$ and $h$ are respectively the electrode surface area and gap
thickness.  We extract the true sample permittivity from the measured
frequency response by performing a complex function fit procedure that
takes into account electrode polarization and capacity---that can be
represented by a constant-phase-angle (``CPA'') element---interfacial
dispersion, also known as the Maxwell-Wagner effect, along with
relaxation processes due to the sample itself. Details about the
deconvolution of the data are given in the Supporting Information. 

We write the measured complex admittance $Y_{m}^{\ast}(\omega)$ as

\begin{equation} 
Y_{m}^{\ast}(\omega)=\left[
  \left(Y^{\ast}(\omega)\right)^{-1}+\left(A(i\omega)^{d}\right)_{CPA}^{-1}
  \right]^{-1}
\end{equation}
where $A$ and $d$ characterize the $\omega$-dependent fractal
polarization due to the blocking electrode, as in \cite{fel98}, and

\begin{equation}
Y^{\ast}(\omega)=\imath\omega\epsilon_{0}\frac{S}{h}\left[
  \epsilon_{\infty}+\sum_{j=1}^\mathcal{N}\frac{\Delta\epsilon_{j}}{\left[1+\left(\imath\omega\tau_{j}\right)^{\alpha_{j}}\right]^{\beta_{j}}}-
\frac{\sigma_{0}}{\imath\omega}\right]
\label{admittance}
\end{equation}
is the admittance of the sample itself---expressed as a conductivity term plus
a combination of Havriliak-Negami functions.  Here $\sigma_{0}$ is the sample
conductivity, $\epsilon_{\infty}$ is the high frequency limit of the
permittivity, $\mathcal{N}$ is the number of relaxation processes
(i.e. $\mathcal{N}=2$ for the broad peak in Fig.~\ref{panel}),
$\Delta\epsilon_{j}$ and $\tau_{j}$ are the dielectric strength and the
relaxation time for the $j$th contribution, respectively, and $\alpha_{j}$ and
$\beta_{j}$ characterize the shape of the relaxation time distribution
function.

The presence of blocking electrodes eliminates the dc-conductivity
across the bulk sample, but not the sample conductivity term
Eq.(\ref{admittance}), related to local displacement of protons along the
protein surface \cite{pag05}.  This sample ``local'' conductivity has
the same temperature dependence as that measured in Ref.~\cite{paw08}.
Since the experimental set-up in Ref.~\cite{paw08} does not use
blocking electrodes, this observation provides a ``second check'' for the
reliability of our data analysis. The check provides also a verification of our data
decomposition, because in Ref.~\cite{paw08} a different 
deconvolution of the data is used, but the results for the same relaxation mode 
are the same.
See the Supporting Information, section {\rm I}, for a
``third check''  of our data
analysis.

\subsection{Quantitative analysis of the shape of the relaxation time
  distribution function}

Raicu \cite{rai99} has proposed a phenomenological ``universal
dielectric response'' function able to describe a single Debye-like
relaxation, such as the Havriliak-Negami (HN) relaxation, combined
with interfacial dispersion and electrode polarization.  An important
result of this work is that a distribution function for the relaxation
times in the frequency domain can be directly calculated using
parameters appearing in the universal response.  We adapt an algorithm
\cite{pro82} to obtain, from raw $\epsilon_{m}^{\ast}(\omega)$ data, a
distribution function in the frequency domain, by means of an inverse
Laplace Transform with no \textit{a priori} assumption on the kind and
number of relaxation processes This approach has proven to be a
reliable tool to obtain a distribution function of relaxation times
\cite{alv91,ban05}. Figure 2 of ref. \cite{bru04} compares the two
distribution functions, one based on a single relaxation, the other
with no assumption on the number of relaxation processes, for the data
set of hydrated lysozyme powder at $270.4$~K and $h=0.26$ g H$_2$O/g dry protein.  The
analysis shows that the distribution function derived for a single
relaxation process plus the effect of electrode polarization and
interfacial dispersion does not totally account for the distribution
calculated with no \textit{a priori} assumptions. This implies that
one or more additional relaxation processes are required to describe
the raw dielectric data. We found that an additional relaxation, is
sufficient to completely account for the calculated distribution
function.  The addition of a third relaxation term results in
unphysical negative $\Delta\epsilon_{i}$, therefore is not considered.
We assume that a relatively small increase of water content from
$0.26$ to $0.3$ g H$_2$O/g dry protein does not change the results of the approach
described above. 

\subsection{Simulations}

We simulate $N=10000$ water molecules in the $NPT$ ensemble. 
To equilibrate such a large system 
at about $180$~K for $100$~sec is a task
that cannot be accomplished with molecular dynamics simulations of 
any detailed model of water. To overcome this problem, we
(i) adopt a coarse-grained model, as described in the main text, and
(ii) use MC simulations. Depending if we want to study
thermodynamic quantities or dynamic quantities, we implement two
different MC techniques (see the Supporting Information, section {\rm II}).

We thank F.~Caupin, H.~Frauenfelder F.~Mallamace, S.~Sastry,
E.~G.~Strekalova, for discussions, NSF grants CHE0616489, CHE0911389,
CHE098218, and CHE0404673 and MICINN-FEDER (Spain) grant FIS2009-10210
for support.



\begin{thebibliography}{100}
\bibitem{nota_MC} To compare with experiments, MC $T$ 
is offset of
  $134.5$~K, MC $P$ is rescaled by $0.507$ and offset of $-2\times
  10^{-4}$~GPa.
\bibitem{KFS} In \cite{KFS1} this crossover was estimated as
VFT
to Arrhenius due to the limited resolution at low $T$.
\end{thebibliography}
\end{document}